\documentstyle[amssymb,12pt]{article}

\begin{document}

\title{G{\" o}del Type  Metrics in Einstein-Aether Theory}
\author{ Metin G{\" u}rses\\
{\small Department of Mathematics, Faculty of Sciences}\\
{\small Bilkent University, 06800 Ankara - Turkey}}

\begin{titlepage}
\maketitle

\begin{abstract}
Aether theory is introduced to implement the violation of the
Lorentz invariance in general relativity. For this purpose a unit
timelike vector field is introduced to the theory in addition to
the metric tensor. Aether theory contains four free parameters
which satisfy some inequalities in order that the theory to be
consistent with the observations. We show that the G{\" o}del type
of metrics of general relativity are also exact solutions of the
Einstein-aether theory. The only field equations are the
3-dimensional Maxwell field equations and the parameters are left
free except $c_{1}-c_{3}=1$.

\end{abstract}
\end{titlepage}


\section{Introduction}

Noncommutativity of local coordinates seems to be an important
implication of  the string theory. Such models are considered in
quantum field theory and it is observed that Lorentz invariance is
broken due to the additional terms coming from the
noncommutativity \cite{kost2},\cite{kost1}. Lorentz violating
theories may lead to some new effects in astrophysics and
cosmology \cite{lim}.

In order to include Lorentz symmetry violating terms in
gravitation theories, apart from some noncommutative gravity
models, one may also consider existence of preferred frames. This
can be achieved admitting  a unit timelike vector field in
addition to the metric tensor of spacetime. Such a timelike vector
implies a preferred direction at each point of spacetime. Here the
unit timelike vector field is called the aether and the theory
coupling the metric and unit timelike vector is called the
Einstein-aether theory.  In the last decade there is an increasing
interest in the aether theory \cite{jac1}-\cite{miya}.

Let $u^{\mu}$ be unit timelike vector ($u^{\mu}\, u_{\mu}=-1$) and
let a four rank tensor $K^{\mu \nu}~_{\alpha \beta}$ be given by

\begin{equation}
K^{\mu \nu}~_{\alpha \beta}=c_{1}\, g^{\mu \nu}\, g_{\alpha
\beta}+c_{2}\, \delta^{\mu}_{\alpha}\,
\delta^{\nu}_{\beta}+c_{3}\, \delta^{\mu}_{\beta}\,
\delta^{\nu}_{\alpha}-c_{4}\, u^{\mu}\, u^{\nu}\, g_{\alpha
\beta},
\end{equation}
where $c_{1}, c_{2}, c_{3}$ and $c_{4}$ are the constants of the
theory. The action of the theory is given as ${\cal L}$
\begin{equation} \label{action}
I={1 \over 16 \pi G}\, \int \,\sqrt{-g}\,{\cal L} \, d^{4}\,x,
\end{equation}
where
\begin{equation}
{\cal L}=R-K^{\alpha \beta}\,\,_{\mu \nu}\, \nabla_{\alpha}\,
u^{\mu}\, \nabla_{\beta}\, u^{\nu}+\lambda\, (u^{\mu}\,
u_{\mu}+1).
\end{equation}

 We define a second rank tensor $J^{\mu}~_{\nu}$ as

\begin{equation}
J^{\mu}~_{\nu}=K^{\mu \alpha}~_{\nu \beta}\, \nabla_{\alpha}\,
u^{\beta}.
\end{equation}
Then the field equations of the aether theory are given by

\begin{eqnarray}
&&G_{\mu \nu}=\nabla_{\alpha}\, [ J^{\alpha}~_{(\mu}
\,u_{\nu)}-J_{(\mu}~^{\alpha}\, u_{\nu)}+J_{(\mu \nu )}\,
u^{\alpha}]+ \nonumber\\
&&+c_{1}(\nabla_{\mu}\, u_{\alpha}\, \nabla_{\nu}\,
u^{\alpha}-\nabla_{\alpha}\, u_{\mu}\, \nabla^{\alpha}\,
u_{\nu}) \nonumber \\
&&+c_{4}\, \dot{u}_{\mu}\, \dot{u}_{\nu}+\lambda u_{\mu}\,
u_{\nu}-{1 \over 2} L \, g_{\mu \nu}, \label{eqn01}\\
&&c_{4}\, \dot{u}^{\alpha}\, \nabla_{\mu}\,
u_{\alpha}+\nabla_{\alpha} \, J^{\alpha}~_{\mu}+\lambda
u_{\mu}=0,\label{eqn02}\\
&&u^{\mu}\, u_{\mu}=-1,
\end{eqnarray}
where $\dot{u}^{\mu}=u^{\alpha}\, \nabla_{\alpha}\, u^{\mu}$ and

\begin{eqnarray}
\lambda&=&c_{4}\, \dot{u}^{\alpha}\, \dot{u}_{\alpha}+u^{\alpha}\,
\nabla_{\beta}\, J^{\beta}~_{\alpha}, \\
L&=&K^{\mu \nu}~_{\alpha \beta}\, (\nabla_{\mu}\, u^{\alpha})\,
(\nabla_{\nu}\, u^{\beta})
\end{eqnarray}

The action given above is invariant under the redefinition
\cite{barb},\cite{foster}

\begin{eqnarray}
\tilde{g}_{\mu \nu}&=&g_{\mu \nu}-(1-B) u_{\mu}\, u_{\nu}, \label{red1}\\
\tilde{u}^{\mu}&=&{1 \over \sqrt{B}} \,u^{\mu}
\end{eqnarray}
where $B$ is a positive constant. Then the parameter transform as

\begin{eqnarray}
c_{1}&=&{1 \over 2B}\,[(1+B^2)\tilde{c}_{1}+(1-B^2)\, \tilde{c}_{3}-1+B^2],\\
c_{2}&=&{1 \over B}\,(\tilde{c}_{2}+1-B),\\
 c_{3}&=&{1 \over
2B}\,[(1-B^2)\tilde{c}_{1}+(1+B^2)\tilde{c}_{3}-1+B^2],\\
c_{4}&=&\tilde{c}_{4}-{1 \over
2B}\,[(1-B^2)\tilde{c}_{1}+(1-B^2)\tilde{c}_{3}\\\nonumber
&&-1+B^2],
\end{eqnarray}

These transformations imply that letting for instance all
$\tilde{c}_{i}, i=1,2,3,4$ to vanish then we have a special type
of aether theory. This implies that  $c_{i}$' s are all related.
$\tilde{g}_{\mu \nu}$ is the metric of a vacuum spacetime and
$g_{\mu \nu}$ is the metric of the aether theory with these
special $c_{i}$' s.

A special case is the Einstein-Maxwell theory with dust
distribution (no pressure) \cite{jac3} (see also \cite{zlos}
 and \cite{bek1}) . Let $c_{2}=c_{4}=0$ and $c_{3}=-c_{1}$. Then
$J^{\mu}\,_{\nu}=c_{1}\, F^{\mu}\,_{\nu}$ and the action above
becomes

\begin{equation}\label{ozel1}
I={1 \over 16 \pi G}\, \int \,\sqrt{-g}\,[R-c_{1}\, F^2+\lambda\,
(u^{\mu}\, u_{\mu}+1)] \, d^{4}\,x
\end{equation}
with the filed equations

\begin{eqnarray}
G_{\mu \nu}=c_{1}\, T_{\mu \nu}+\lambda u_{\mu}\,u_{\nu}, \label{ozel2}\\
\nabla_{\mu}\, F^{\mu \nu}= {\lambda \over c_{1}} \, u^{\nu},\label{ozel3}\\
u^{\mu}\, u_{\mu}=-1, \label{son}
\end{eqnarray}
where $F_{\mu \nu}=\nabla_{\nu}\, u_{\mu}-\nabla_{\mu}\, u_{\nu}$
and $T_{\mu \nu}$ is the energy momentum tensor of the filed
$F_{\mu \nu}$. This theory differs from the Einstein Maxwell
theory due to the last equation (\ref{son}) which brakes the gauge
invariance of the theory. A generalization of the above special
theory is given in \cite{bek1}, called TeVsS. This theory contains
also a scalar (dilaton) field coupling to the unit timelike vector
field and the metric tensor.

 The parameters $c_{1}, c_{2}, c_{3}$ and $c_{4}$
are not so free. They satisfy some inequalities in order that the
aether theory to be compatible with some observations \cite{jac2},
\cite{miya}.

\vspace{0.3cm}

\noindent 1. It is shown that this theory has the same PPN
parameters as those of general relativity if

\begin{equation}
c_{2}={-2c_{1}^2-c_{1}c_{3}+c_{3}^2 \over 3c_{1}},~~
c_{4}=-{c_{3}^2 \over c_{1}}
\end{equation}

\vspace{0.3cm}

\noindent
 2. In the slow motion limit of the theory the constant
playing the role of  Newton's constant is

\begin{equation}
G_{N}=G\, (1-{c_{1}+c_{4} \over 2})^{-1}
\end{equation}
where $G$ is  Newton's constant.

\vspace{0.3cm}

\noindent 3. In Friedman-Robertson-Walker type of cosmological
models, the theory admits the cosmological gravitational constant

\begin{equation}
G_{cosmos}=G\, (1+{c_{+}+3 c_{2} \over 2})^{-1}
\end{equation}
where $c_{+}=c_{1}+c_{3}$.

\vspace{0.3cm}

\noindent 4. Primordial abundance of $^{4}\mbox{He}$ gives

\begin{equation}
|{G_{cosmo} \over G_{N}}-1| < {1 \over 8}
\end{equation}

\vspace{0.3cm}

\noindent 5. From the maximum mass of the neutron stars

\begin{equation}
c_{1}+c_{4} \le 0.5 - 1.6
\end{equation}

\vspace{0.3cm}

\noindent 6. Stability against linear perturbations in Minkowski
background we have

\begin{equation}
0<c_{1}+c_{3} <1, ~~ ~0 < c_{1}-c_{3}< {c_{+} \over 3(1-c_{+})}
\end{equation}

\vspace{0.3cm}

\noindent For example, when $c_{1}-c_{3}=1$ the above constraints
are all satisfied where ${G_{cosmo} \over G_{N}}=1$, and ${7 \over
8} < c_{1} <1$ , $-{1 \over 8} <c_{3} <0$.

In this work we first give the G{\" o}del type metrics in general
relativity by presenting a short summary of  \cite{gur1}. We show
that the G{\" o}del type metrics form an exact solution of the
Einstein field equations with a charged dust distribution. The
only field equations to be solved are the three dimensional
Euclidean Maxwell equations. Next we show that G{\" o}del type
metrics solve also the field equations of the Einstein-aether
theory. The only remaining equations are again the three
dimensional Euclidean Maxwell equations corresponding the unit
timelike vector field and the constraint $c_{1}-c_{3}=1$. It seems
that this  constraint is compatible with the bounds of the
parameters of Einstein-aether theory.

\section{G{\" o}del Type Metrics in General Relativity}

Let $u^{\mu}=-\delta^{\mu}_{0}$ be a timelike vector with
$u_{0}=1$  in four dimensional spacetime $M$ and $h_{\mu \nu}$ be
a constant tensor ($\partial_{\alpha}\, h_{\mu \nu}=0$) such that
$u^{\mu}\, h_{\mu \nu}=0$. G{\" o}del type of metrics are defined
by \cite{gur1}

\begin{equation}\label{metric}
g_{\mu \nu}=h_{\mu \nu}-u_{\mu}\, u_{\nu}
\end{equation}
It is easy to show that $u^{\mu}$ is also a Killing vector of the
spacetime geometry ($M,g$). Then we can define an antisymmetric
tensor $f_{\mu \nu}$ as

\begin{equation}
f_{\alpha \beta}=u_{\beta, \alpha}-u_{\alpha,\beta}=2 u_{\beta;
\alpha}
\end{equation}
where semicolon denotes covariant derivative with respect to the
Christoffel symbol. The Christoffel symbol corresponding to the
metric (\ref{metric}) is

\begin{equation}
\Gamma^{\mu}_{\alpha \beta}={1 \over 2}\,(u_{\alpha}\,
f^{\mu}\,_{\beta}+u_{\beta}\, f^{\mu}\,_{\alpha})-{1 \over
2}(u_{\alpha, \beta}+u_{\beta, \alpha})\,u^{\mu}.
\end{equation}
It is easy to show that

\begin{equation}
u^{\alpha}\, \partial_{\alpha}\, u_{\beta}=0,~~~ u^{\alpha}\,
f_{\alpha \beta}=0.
\end{equation}
Then

\begin{equation}
\dot{u}^{\mu}=u^{\alpha}\, u^{\mu}\,_;{\alpha}=0
\end{equation}
It is now straightforward to show that the Einstein tensor becomes

\begin{equation}\label{god01}
G_{\mu \nu}={1 \over 2}\,T^{f}_{\mu \nu}+{1 \over 4} f^{2}\,
u_{\mu}\,u_{\nu}
\end{equation}
provided $f$ satisfies the equation

\begin{equation}\label{maxeqn}
\partial_{\alpha}\, f_{\mu}\,^{\alpha}=0
\end{equation}
where $T^{f}_{\mu \nu}$ is the Maxwell energy momentum tensor for
the antisymmetric tensor $f_{\mu \nu}$
\[
T^{f}_{\mu \nu}=f_{\mu \alpha}\,f_{\nu}\,^{\alpha}-{1 \over 4}\,
f^2\, g_{\mu \nu}
\]
where $f^2=f^{\alpha \beta}\, f_{\alpha \beta}$. Maxwell's
equations (\ref{maxeqn}) can also be written as

\begin{equation}\label{maxeqn1}
\nabla_{\alpha}\, f^{\alpha \mu}={1 \over 2}\, f^2\, u^{\mu}
\end{equation}
Hence G{\" o}del type metrics (\ref{metric}) satisfy the Einstein
field equations with charged dust distributions. The only field
equations are the  Maxwell equations (\ref{maxeqn}) or
(\ref{maxeqn1}) which may further be reduced to a more simpler
form

\begin{equation}\label{maxeqn2}
\partial_{i}\, f_{ij}=0
\end{equation}
Hence there is no electric field ($u^{\mu}\, f_{\mu i}=f_{0i}=0$),
only the magnetic field exists. Maxwell equations (\ref{maxeqn2})
are in Euclidean 3-dimensions. We exhibited some solutions of this
equation and hence explicit G{\" o}del type metrics in
\cite{gur1}. All such spacetimes contain closed timelike and
closed null curves. Some examples of these metrics are are given
as follows \cite{gur1}:

\vspace{0.5cm}

\noindent {\bf a)}. Let $u_{\mu}\, dx^{\mu}=dt+b\,(x^2\,
dx^{1}-x^{1}\, dx^{2})$. Then $f_{ij}\, dx^{i}\wedge dx^{j}=2b\,
dx^{1}\,\wedge dx^{2}$. Hence (\ref{maxeqn2}) is satisfied
identically. The the metric and the unit timelike vector $u_{\mu}$
in cylindrical coordinates are given by

\begin{eqnarray}
ds^2&=&-(dt-b\rho^2 d\phi)^2+d\rho^2+\rho^2\, d\phi^2+dz^2,\\
u_{\mu}\, dx^{\mu}&=&dt-b \rho^2\, d\phi,
\end{eqnarray}
where $b$ is an arbitrary constant.

\vspace{0.5cm}

\noindent {\bf b)}. Let $u_{i}=\psi\, \delta^{3}_{i}$ where $\psi$
is an harmonic  function of $x^1 $ and $ x^2$ ($\nabla^2\,
\psi=0$). Then the Maxwell equations (\ref{maxeqn2}) are satisfied
identically. Metric tensor and the unit timelike vector field
$u_{\mu}$ are given by

\begin{eqnarray}
ds^2&=&-(dt-\psi(\rho,\phi) dz)^2+d\rho^2+\rho^2\, d\phi^2+dz^2,\\
u_{\mu}\, dx^{\mu}&=&dt+\psi(\rho,\phi) dz.
\end{eqnarray}

We have also considered  the G{\" o}del type metrics when $u_{0}$
is not a constant \cite{gur2}. In this case the proposed metric
constitute  an exact solution of several theories with a dilaton
field.

\section{G{\" o}del Type Metrics in Aether Theory}

We assume that the metric $g_{\mu \nu}$ and the timelike four
vector $u^{\mu}$ are the G{\" o}del type metric and the timelike
vector defined in the previous section. With these assumptions we
find that

\begin{eqnarray}
J^{\mu}\,_{\nu}&=&{1 \over 2}\, (c_{1}-c_{3})\, f^{\mu}\,_{\nu}, \label{eqn03}\\
\lambda&=&-{1 \over 4}\, (c_{1}-c_{3})\, f^2, \label{eqn04}\\
L&=&{1 \over 4}\, (c_{1}-c_{3})\, f^2 \label{eqn05}
\end{eqnarray}
Then the Einstein field equations (\ref{eqn01}) becomes

\begin{equation}
G_{\mu \nu}=(c_{1}-c_{3})\,[{1 \over 2}\,T^{f}_{\mu \nu}+{1 \over
4} f^{2}\, u_{\mu}\,u_{\nu}]
\end{equation}
and the aether equation (\ref{eqn02}) reduces to

\begin{equation}\label{maxeqn2}
\nabla_{\alpha}\, f^{\alpha \mu}={1 \over 2}\,(c_{1}-c_{3})\,
f^2\, u^{\mu}
\end{equation}

\noindent Comparing these with (\ref{god01}) and (\ref{maxeqn1})
we get $c_{1}-c_{3}=1$. Hence the only field equations remaining
for the Einstein-aether theory are those given in (\ref{maxeqn2}).
This result shows that the Einstein-aether and the Einstein
theories are equivalent under the assumptions when the metric is
the G{\" o}del type and $c_{1}-c_{3}=1$. In both cases the
spacetime is curved due to the unit timelike vector field and the
matter distribution is a charged dust due to the same unit
timelike vector field.

\vspace{0.5cm} \noindent G{\" o}del type metrics solve also the
special case of the aether theory (charged dust case) given in
(\ref{ozel1})-(\ref{ozel3}) but in our case the parameters $c_{2}$
and $c_{4}$ are not necessarily zero and $c_{3} \ne -c_{1}$. We
find $c_{1}={1 \over 2}$ and $\lambda={f^2 \over 4}$. Here we take
the zeroth component of the vector field  as unity, $u_{0}=1$.
When we relax this condition  one needs to introduce a scalar
field into the theory. Such a theory, called TeVeS, is given in
\cite{bek1}. We conjecture that  G{\" o}del type metrics with
nonconstant $u_{0}$ form a class of exact solutions of TeVeS.

\vspace{0.3cm}

\noindent
 Here we have some remarks: (1).\, G{\" o}del type metrics (\ref{metric})
 we defined here and in
the previous section differ from the metric redefinition in
(\ref{red1}) because $h_{\alpha \beta}$ in G{\" o}del type metrics
is a degenerate matrix, its determinant is equal to zero. Hence
the G{\" o}del type metrics form a distinct class of exact
solutions of the aether theory. (2).  Although there are no closed
timelike or closed null geodesics in G{\" o}del type of
spacetimes, they contain closed null worldlines \cite{gur3}. It
seems that violation of Lorentz invariance implements  causality
violations.

\section{Conclusion}

In \cite{gur1} and \cite{gur2} we have shown that
 G{\" o}del type metrics arise in several low energy limits of
 string theory . In all these theories we have reduced the field
 equations to Maxwell type of equations in
 various dimensions. Several exact solutions with their properties were
 exhibited. Among the properties of these solutions we can mention
 the existence of closed timelike and closed null curves. In this
 work we carried these solutions,  the G{\" o}del type metrics, to
 the Einstein-aether theory. We proved that the  G{\" o}del type metrics
 reduce the complete field equations of the theory to three
 dimensional Maxwell equations corresponding to the unit timelike
 vector field where all parameters of the theory are left free except  $c_{1}-c_{3}=1$.
  We also showed that G{\" o}del type
 metrics solve  a special reduction of the aether theory
 \cite{jac2}, \cite{zlos}, \cite{bek1}.

 The G{\" o}del type metrics used in this work has $u_{0}=1$.
 If we relax this condition, these metrics constitute a class of exact solutions
 of several low energy limits of string theories in various
 dimensions with nonconstant dilaton field. We claim that
 the G{\" o}del type metrics with nonconstant $u_{0}$ solve field equations of an aether
 theory, like TeVeS \cite{bek1} or its modifications,  with a dilaton field.

 In aether theories, in addition to the spacetime metric a unit
 timelike vector field is considered which implies existence of
 preferred unit timelike direction at each point of spacetime,
 breaking the local Lorentz symmetry. An
 alternative to the timelike vector field we may  consider
  a dynamical null vector. The action will be similar to the one
  given in (\ref{action}) except the coefficient of the $\lambda$
  term. We conjecture that
  Kerr-Schild metrics will form a class of exact solutions of such
  theories. All these issues will communicated later.

 \vspace{2cm}
 This work is partially supported by the Scientific and
Technological Research Council of Turkey (TUBITAK) and  Turkish
Academy of Sciences (TUBA).

\end{document}